\documentclass[12pt]{article}

\usepackage{graphicx}
\usepackage{cite}

\begin{document}

\begin{center}

   {\Large \bf Modelling the onset of oxide formation \\ 
               on metal surfaces from first principles}

   \vspace{0.5cm}  Lucio Colombi Ciacchi

   \vspace{0.25cm}{\it
       Fraunhofer Institut f\"ur Werkstoffmechanik IWM,\\
       Freiburg, Germany, and} \\
   \vspace{0.15cm}{\it
       Institut f\"ur Zuverl\"assigkeit von Bauteilen und Systemen, \\
       Universit\"at Karlsruhe, Karlsruhe, Germany.} \\

\end{center}

\vspace{0.5cm}
The formation of ultrathin oxide layers on metal surfaces is a 
non-thermally-activated process which takes place spontaneously at very
low temperatures within nanoseconds.
This paper reports mechanistic details of the initial oxidation of
bare metal surfaces, in particular Al(111) and TiN(001), as obtained
by means of first-principles molecular dynamics modelling within the
Density-Functional Theory.
It is shown that the reactions of bare metal surfaces with O$_2$
molecules take place according to a ``hot-atom'' dissociative
mechanism which is triggered by the filling of the $\sigma^{\ast}$
antibonding molecular orbital and is characterised by a sudden 
release of a large amount of kinetic energy.
This released energy provides a driving force for metal/oxygen
place-exchange processes which are responsible for the onset of oxide
formation at virtually 0~K and at oxygen coverages well below 
1~monolayer (ML).
Further simulations of the oxidation reactions reveal that a
disordered ultrathin oxide forms on Al(111), whereas a rather ordered
structure develops on TiN(001) following a selective oxidation process
which leaves clusters of Ti vacancies in the TiN lattice underneath
the oxide layer.

\newpage

\section*{1. Introduction}

Metal surfaces in contact with the atmosphere become quiclky covered 
by oxide layers of thicknesses which vary from few atomic layers 
(less than 1~nm) to several microns, depending on the reactivity of 
the metal and the history of the sample~\cite{Lawless_1974}.
Even the noblest metals, such as gold, may chemically interact with 
oxygen species~\cite{Hua_2006} at their surfaces.
The oxide skin governs the chemical and physical interactions of the
metal with the outer environment, thus a precise knowledge of its
structure and composition is fundamental to a number of
technologically important problems.
These comprise for instance the behaviour of metallic biomedical
implants within the chemically aggressive physiological
environment~\cite{Piscanec_2004}, the response of electrodes or
sensors in direct contact with liquid solutions~%
\cite{Jerkiewicz_2004,Diniz_2004}, the performances of metallic
contact structures in electronic devices, the wear and friction
properties of metal surfaces~\cite{Fehlner_1986}, the rate of
sub-critical crack propagation due to stress-corrosion
phenomena~\cite{Muhlstein_2002}, and many others~\cite{Fromm_1998}.

Traditionally, engineers have been much concerned with oxidation
phenomena at high temperatures, since they result in a continuous
consumption of metal and therefore inevitably limit the life of
machine components such as, for instance, turbine blades~%
\cite{Lawless_1974}.
For this reason, great effort has been expended in the last century 
to develop a basic understanding of the kinetic laws governing
high-temperature oxidation, and several theories have been formulated
which explain and predict the oxidation behaviour of many metals and
metal alloys~\cite{Tammann_1920,Cabrera_1948,Fehlner_1970}.
However, as noted in~\cite{Lawless_1974}, many of the existing rate
laws seem to be satisfactory only over a limited oxide thickness
range, and often are not transferable to low temperature or low
pressure regimes.

In particular, detailed knowledge of processes of oxide formation at
room temperature or below is scarce compared with the amount of
information available in the field of high-temperature oxidation.
The available kinetic models rely on assumptions on the nature of
characteristic defects in the formed oxide, which govern the diffusion
processes necessary to oxide growth~\cite{Jeurgens_2002,Reichel_2006}.
In turn, this requires a precise knowledge of the structure and
composition of the ultrathin oxide layers which form spontaneously
when a bare surface is put in contact with an oxidising environment.
These layers are often glassy oxides of non-integer stoichiometry,
so structural information on the oxide lattice or on the defects 
present cannot be inferred from the corresponding bulk oxide
structures which form at more elevated temperatures or oxygen
pressures~\cite{Lundgren_2002}.
Moreover, in the case of ultrathin oxide layers it is very difficult, 
if not impossible, define a precise metal/oxide interface
as in the case of thicker crytalline oxides grown on 
metals (see e.g.~\cite{Benedek_1999,Gemming_2001}).

Differently to the growth of oxide layers at high temperature, which
is limited basically by the thermally activated diffusion of ionic
species across the oxide layer, the initial reactions of a bare metal
surface with an oxidising atmosphere are not thermally activated.
Namely, spontaneous formation of a few monolayers of oxide
has been observed at temperatures as low as a few K~%
\cite{Ruckman_1992,Jacobsen_1995}.
Formation of an oxide layer takes place if adsorbed oxygen atoms can 
change place with underlying metal atoms and be incorporated below
the metal surface.
Driving forces for this metal/oxygen place-exchange process are
in principle available due to the generally high heat of adsorption
of oxygen on metals, and due to the large electric field which
develops at the metal/oxide interface as a consequence of electron
donation from the metal to the incoming oxygen~%
\cite{Cabrera_1948,Fehlner_1970}.
However, the mechanistic details of the actual exchange process 
are still unknown, and it remains to be understood how this can
take place at virtually 0~K~\cite{Fehlner_1986}.

Our goal is  to gain a mechanistic insight into the events of
initial oxide formation on metal surfaces, prior to the beginning
of a diffusion-limited oxide growth process.
As these events are not thermally activated and occur on a scale of
picoseconds to nanoseconds, they are hardly accessible through
commonly available experimental techniques.
Therefore, they will be addressed in this work by means of
``first-principles'' atomistic modelling at the quantum mechanical
level~\cite{Payne_1992}.
Helped by the fast development of accurate and efficient theoretical
formalisms and by the ever-increasing power of computer platforms,
this is becoming one of the methods of choice to investigate chemical
processes in materials science at the atomic scale~\cite{Hafner_2000}.
In particular, so-called first-principles thermodynamics approaches
have been widely used in the last few years to address the structure
and composition of thin oxide layers on metal surfaces 
(see e.g.~\cite{Lundgren_2002,Reuter_2003,Finnis_2005}).
Such approaches give precise information on the different oxide phases
and structures which are thermodynamically stable at various
temperatures and oxygen pressures.
However, low-temperature oxidation reactions may lead to formation
of intrinsically metastable, often amorphous, oxide structures.
Therefore, here we will investigate their formation via
a first-principles molecular dynamics approach.
This technique is obviously limited by the small system size and the
short simulation time addressable, but provide us with a detailed
information on the dynamical features of the oxidation reactions which
would be very difficult to gain otherwise.
Moreover, using molecular dynamics we are able to obtain structural
models of metastable glassy oxide layers which may capture the
physical and chemical properties of ultrathin oxide layers formed at
low temperatures more realistically than the corresponding
thermodynamically stable phases.
The growth of oxide scales of larger thickness can be simulated
using atomistic techniques at the classical level, which, however, 
necessarily need to be tuned on previous knowledge from unbiased 
simulations at the quantum level~\cite{Vashishta_1999,Vashishta_2006}.

\section*{2. Computational details}

All the results presented in the following sections were obtained
using molecular dynamics techniques in which the forces acting on the
atoms are computed within the Density-Functional Theory (DFT) quantum
mechanical formalism.
Using a first-principles approach, as comprehensively reviewed in~%
\cite{Payne_1992,Hafner_2000,Galli_1993,Marx_2000},
is strictly necessary to investigate processes involving creation
and rupture of covalent bonds.
After computing the forces within the DFT, the atoms are moved
according to the classical equation of motion integrated with
standard algorithms~\cite{Verlet_1967}.
All the calculations in the present work were carried out in this
theoretical framework, using the {\sc Lautrec}
code~\cite{Lautrec}.
Both the minimisation of the electronic states and the dynamics
of the atoms were performed using the Car--Parrinello (CP)
method~\cite{Car_1985} and the gradient-corrected exchange-correlation
potential PW91~\cite{Perdew_1992}.
Either separable, norm-conserving atomic
pseudopotentials~\cite{Troullier_1993} or Bl\"ochl's Projector
Augmented Waves approach~\cite{Blochl_1994} have been used to describe
the electron--nucleus interactions including in the simulations only the
valence electrons.
The CP simulations of metallic systems were performed with the algorithm
proposed in Refs.~\cite{VandeVondele_1999} and~\cite{Stengel_2000}.
The simulated systems are periodically repeated in all directions
of space.
Surface models were constructed including in the simulation cell a
slab of a few layers of metal atoms separated from its periodically
repeated image by a vacuum layer in the direction perpendicular to the
surface.
The thickness of the slab and of the vacuum layer have been chosen
so as to avoid spurious effects due to truncation of the crystal
or interactions with the image systems.
Details of the simulation parameters and test calculations to assess the
precision of the formalism can be found, in particular, 
in~\cite{Ciacchi_2004,Ciacchi_2005,Piscanec_2004}.

\section*{3. O$_2$ adsorption on bare metal surfaces}

\subsection*{\rm 3.1. Experimental background}

Bare metal or semiconductor surfaces are intrinsically very
reactive systems.
The cleavage of a crystal along a given crystallographic direction
leaves under-coordinated surface atoms and broken, ``dangling''
bonds which readily participate in chemical reactions with
molecules in proximity of the surface.
A pictorial representation of the dangling bonds of a Si(001)
surface is given in Fig.~\ref{f:si_001_surf} (top), where
the computed Electron Localisation Function (ELF)~\cite{Becke_1990} 
of the system is depicted as a green isosurface.
Regions of large ELF far from the atomic cores and of chemical
bonds are indicative of the presence of lone electrons, and thus
of potentially reactive sites for chemical reactions.
In fact, any Si surface put in contact with the atmosphere 
at normal temperature and pressure conditions would promptly 
react with dioxygen and water molecules and become covered by 
an amorphous ultrathin hydroxylated oxide layer within a few 
nanoseconds (Fig.~\ref{f:si_001_surf}, bottom).
Metal surfaces behave in the same way, although the electrons
available for donation into oxidising molecules are rather 
uniformly delocalised over the atoms composing the surface.

The reactions between O$_2$ molecules and bare metal surfaces have
been studied using scanning tunneling microscopy (STM) under ultrahigh
vacuum conditions~\cite{Brune_1992,Schmid_2001,Komrowski_2001}.
When a previously cleaned metal surface is treated with a flow 
of oxygen gas at low pressure, pairs of oxygen atoms separated
by distances of about 1 to 3 nearest-neighbour metal--metal distances
become visible under the STM (Fig.~\ref{f:al_o2_stm}).
This suggests that O$_2$ reacts dissociatively with metal surfaces 
by cleavage of the O--O bond, leading to pairs of single adsorbed 
oxygen atoms.
On the Al(111) surface at room temperature or below, the reaction may
proceed with adsorption of only one O atom and ejection of the second
O atom in the gas-phase~\cite{Komrowski_2001}.
This process is still thermodynamically favoured since the heat of
adsorption of just a single atom on Al(111) is more than 7~eV, while
the energy required to break the O--O bond of O$_2$ is slightly larger
than 5~eV.
Both the possibility of O ejection and the large distances
(up to $\sim$10~\AA) which separate pairs of atoms after
the oxidation reaction at very low temperatures indicate that
the dissociation process of O$_2$ on the metal surface
is associated with a sudden release of high kinetic
energy.
This energy release enables the atoms to overcome relatively large
diffusion barriers over the surface (of the order of $\sim$0.5~eV)
and to cover distances much larger than a normal thermally activated 
diffusion process would allow at the low reaction temperatures.
Therefore, the term ``hot-atom''-mechanism has been used
in the literature to describe the non-thermally-activated
process of dioxygen dissociation on metals~%
\cite{Brune_1992,Wintterlin_1996}.

In the following section, first-principles molecular dynamics (FPMD)
simulations of the process of dioxygen dissociation on several metal
surfaces will be reported, focusing on the actual driving force for
the occurring ``hot-atom'' dissociation event.

\subsection*{\rm 3.2. ``Hot-atom'' dissociation of O$_2$ on bare metal surfaces}

Before describing the results of FPMD simulations of the
dissociative adsorption of dioxygen on metallic surfaces, it
is useful to summarise very briefly some features of the 
molecular orbitals of O$_2$ (Fig.~\ref{f:O2_orbitale}).
Dioxygen is a paramagnetic molecules, having the two 
degenerate $\pi^{\ast}$ antibonding orbitals each filled 
with one electron of equal spin.
The empty $\sigma^{\ast}$ orbital lies almost 7~eV higher 
in energy than the $\pi^{\ast}$ orbitals, and the p$_x$
orbital lobes along the direction of the O--O bond are
evidently compressed in the interatomic bond region 
(see Fig.~\ref{f:O2_orbitale}, right).
It needs to be noted that within our ground state DFT method we are
not able to address excited electronic states of dioxygen, which
may in principles have a role in the oxidation reactions.
However, as discussed in ~\cite{Ciacchi_2004}, the mechanisms
of onset of oxide formation are expected not to be influenced
qualitatively by the adiabaticity of the DFT simulations.

The adsorption of an O$_{2}$ molecule on the Al(111) surface has
been simulated using a periodically repeated surface slab of five 
(111) atomic layers separated by a vacuum layer of the same thickness 
(Fig.~\ref{f:al_dyn_2}).
When an oxygen molecule is placed near this surface slab, the two
unpaired electrons in the $\pi^{\ast}$ molecular orbitals can
be visualised by imaging of the spin-density of the system.
This is defined as the difference $\Delta\rho$ between the
particle densities $\rho_{\alpha}$ and $\rho_{\beta}$ associated
with each spin-manifold.
In Fig.~\ref{f:al_dyn_2}a, the O--O axis of the oxygen molecule is perpendicular
to the plane of the page, and the radial symmetry of the $\pi^{\ast}$ 
antibonding orbitals is evident.
When the molecule is about 3.0~\AA\ above the surface, the integrated 
spin density is about 1.8 electrons, indicating a partial charge 
donation from the surface into the molecule, and thus a chemical interaction 
between molecule and surface.
This is also revealed by analysis of the local density of states (LDOS)
integrated within spheres around the O atoms with radii corresponding
to the covalent radius of oxygen (Fig.~\ref{f:graph_o2_fcc_all}).
Namely, the lower edge of the LUMO orbital of dioxygen is placed
at the Fermi level of the system (Fig.~\ref{f:graph_o2_fcc_all}, top)
so that the orbital is not completely empty.
As a general shortcoming of standard DFT techniques, the barrier for
passing from a physisorbed molecular state to a chemically adsorbed
(and dissociated) state on the Al surface is missing (for a thorough
discussion of this issue, see e.g.~\cite{Yourd_2002}).

Indeed, after starting the dynamics, the molecule is quickly
adsorbed on the surface in a process which is characterised by a
full quenching of the spin density due to electron donation 
into the half-filled $\pi^{*}$ antibonding orbital of O$_{2}$ 
(Fig.~\ref{f:al_dyn_2}b, c and Fig.~\ref{f:graph_o2_fcc_all}, middle).
This leads to a gradual increase of the O-O distance, until also the
$\sigma^{*}$ antibonding orbital of the adsorbed molecule becomes
partially populated, as revealed by analysis of the spin density 
(Fig.~\ref{f:al_dyn_2}d) and of the LDOS 
(Fig.~\ref{f:graph_o2_fcc_all}, bottom).
The O--O bond dissociates as a consequence of this event, the O--O
distance increasing abruptly  in a way which is indicative of a
non-thermal, ``hot-atom'' mechanism (Fig.~\ref{f:al_dyn_2}e, f).
The sudden release of kinetic energy during the dissociation 
is large enough to push one of the O atoms below the Al surface,
while one Al atom is pulled out of the surface (Fig.~\ref{f:al_dyn_2}e).
However, in the subsequent dynamics the O atom emerges again 
above the surface and the gained kinetic energy is gradually 
transferred via the lattice vibrations to the whole surface slab.
After quenching of the atomic motion, the O atoms are stably
adsorbed in hollow surface sites, separated by a distance of
5.9~\AA, i.e.\ by more than two Al--Al distances.

In a number of similar FPMD simulations, the same ``hot-atom'' 
dissociation mechanism has been observed to lead to spontaneous 
O$_2$ dissociation on Al(100), Ti(0001), Co(0001), Cr(110), 
TiN(001), and notably also on Si(001)~\cite{Ciacchi_2005}.
We may thus consider it as a general mechanism for the initial
reaction of dioxygen molecules with bare metal or semiconductor
surfaces.
Among the system investigated so far, only the reaction of dioxygen
with Pt(111) resulted in stable adsorption of molecular O$_2$ 
without dissociation.
In fact, the experimental evidence is that dissociation of O$_2$ on
Pt(111) occurs only at temperatures higher than
$\sim$150~K~\cite{Wintterlin_1996}, while our simulations are all
performed starting from fully relaxed systems, i.e.\ at virtually 0~K.
However, when dissociation is observed experimentally, then the
process bears all the typical features of a ``hot-atom''
mechanism~\cite{Wintterlin_1996}.
This suggests that thermal motion of the surface is necessary to
activate the O$_2$ molecule so that the O--O distance becomes
sufficiently large to enable partial filling of the $\sigma^{\ast}$
antibonding molecular orbital.
After that, we expect the dissociation to proceed similarly
to all other cases considered.

The dynamical evolution of an O$_2$ molecule during a FPMD
simulation of its adsorption and dissociation on the Al(100) 
surface is reported in Fig.~\ref{f:graph_o2_100_dist_char}.
In the top graph the evolution of the intramolecular O--O distance 
is plotted together with the evolution of the  distance between 
the centre of mass of the two O atoms and the surface.
The bottom graph reports the evolution of the spin density and the
charge on the atoms integrated within the Bader
regions~\cite{Bader_1990} associated with both O atoms.
It is interesting to note that after the initial adsorption of 
the molecule on the surface (left vertical dotted line), the
charge on the molecule is only slightly more than 1.0 electron.
Only as a consequence of the ``hot-atom'' dissociation event
(right vertical dotted line), which suddenly cleaves the
molecular bond, do the O atoms become charged each with almost
two electrons.
This highlights the fact that the hyperthermal dissociation is 
not triggered by Coulomb repulsion between the O atoms, but 
by the Pauli repulsion between the partially occupied p-type 
orbitals along the O--O axis~\cite{Ciacchi_2004}.

\section*{4. Spontaneous formation of ultrathin oxide layers}

\subsection*{\rm 4.1. Onset of oxide formation at increasing oxygen coverages}

The same technique used to simulate the adsorption of
single oxygen molecules on bare surfaces has been applied
to investigate further oxidation reactions at increasing
coverages of adsorbed oxygen on the Al(111), Ti(0001) and
TiN(001) surfaces.
Namely, for each system, consecutive FPMD simulations of the
adsorption of oxygen molecules have been performed starting with
a new O$_2$ molecule about 3.0~\AA\ above the relaxed surface 
obtained in the previous simulation.
On both Al(111) and Ti(0001), O$_2$ is again observed to adsorb and
dissociate spontaneously with a sudden release of kinetic energy.
Interestingly, in both systems the process lead to incorporation
of O atoms underneath the surface layer at a coverage of 0.50~ML.
At this coverage, on Al(111) an Al vacancy is created temporarily 
as a consequence of the ``hot-atom'' O$_2$ dissociation, and is
quickly filled by a previously adsorbed O atom nearby
(Fig.~\ref{f:o2_03_surfs}, left).
This process represents the onset of formation of an oxide phase, 
as pointed out in Ref.~\cite{Brune_1993}.
The fact that this is observed in our simulation at a coverage well 
below the saturation of one (1$\times$1) O adlayer is indeed
consistent with the experimental finding that oxide nucleation 
can start at coverages as low as 0.2~ML~\cite{Brune_1993}.
A similar scenario is observed on Ti(0001), where the motion of the
surface Ti atoms induced by the energy released upon dissociation 
is such that a previously adsorbed O atom is able to penetrate the 
surface layer and bind to the Ti layer underneath 
(Fig.~\ref{f:o2_03_surfs}, middle).
These simulations strongly suggest that place-exchange processes 
between metal atoms and O atoms are spontaneously activated 
during the oxidation reaction due to the high kinetic energy 
associated with the cleavage of the O--O bond of incoming O$_2$
molecules.

Interestingly, such place-exchange mechanisms are not observed
in simulations of oxidation reactions on the TiN(001) surface
at low oxygen coverages.
In this case, no O atoms are incorporated underneath the surface, 
whereas Ti atoms are observed to escape from the topmost surface 
layer (Fig.~\ref{f:o2_03_surfs}, right) and bind to the
incoming oxygen.
Also, not all O$_2$ molecules are observed to dissociate
spontaneously, indicating a decreased reactivity of the 
partially oxidised surface with respect to the bare TiN surface.
In particular, increasing the simulation temperature to 600 K 
was not sufficient to dissociate the molecule depicted in 
Fig.~\ref{f:o2_03_surfs}(right) within 2.5~ps of simulation
time.

\subsection*{\rm 4.2. Formation of ultrathin oxide layers}

Further FPMD simulations of reactions with O$_2$ molecules with the
partially oxidised Al(111) and TiN(001) surfaces were performed to
increase the oxygen coverage beyond the onset of oxide formation
described in the previous section.
These are described separately for each system in the following
two subsections.

\subsubsection*{\rm Formation of amorphous oxide on Al(111)}

On Al(111) further reactions with dioxygen proceeded spontaneously at
virtually 0~K (i.e., both the surface and the molecules were fully
relaxed in the initial configuration of each FPMD run) up to a
coverage of 1.0~ML~\cite{Ciacchi_2004}.
After that, O$_2$ molecules initially placed at a distance of about
3.0~\AA\ above the surface were repelled by the already formed
ultrathin oxide structure.
Therefore, O$_2$ molecules were placed in close contact with Al atoms
at potentially reactive sites on the outer oxide surface (the initial
Al--O distances were about 2.0 to 2.5~\AA).
Although some of the molecules were again repelled by the surface and
desorbed, in a few cases spontaneous dissociation took place, leading
to growth of the oxide layer up to a coverage of 1.5~ML.
The final structure obtained is reported in Fig.~\ref{f:oxide_1.5ML_al_111}.
Noteworthy is the evident disorder in the oxide structure along the
direction perpendicular to the surface, while a top-view (bottom
left in Fig.~\ref{f:oxide_1.5ML_al_111}) reveals a certain order 
in the surface plane, in which the O atoms are arranged in positions
roughly correspondent to the hollow sites of the Al(111) plane.
Also interesting is the fact that the initially formed oxide structure
presents large cavities at the metal/oxide interface, which seem to be
consistent with the low density of the thin amorphous oxide on Al(111)
observed in x-ray photoelectron spectroscopy (XPS)
experiments~\cite{Snijders_2005}.
In fact, the XPS evidence suggests that only after reaching a critical
thickness does the oxide layer undergo a transition to a crystalline
state, which is associated with a densification of the oxide layer~%
\cite{Snijders_2005}.

It needs to be stressed that the dynamics of the spontaneous reactions 
described above, and in particular the occurrence of metal/oxygen
place-exchange processes after dissociative adsorption of O$_2$
molecules, does lead to oxide structures which are not necessarily
in thermodynamical equilibrium.
This is evident by looking at the computed total energy per unit cell
of the oxidised Al(111) surface at an oxygen coverage of 1~ML (the
unit cell contains 60 Al atoms and 12 O atoms).
The energy of the disordered structure obtained after consecutive
FPMD simulations is $\sim$3~eV higher than the energy of an
ordered monolayer of adsorbed O atoms on fcc surface sites.
However, the energy released after the oxidation reactions is very
quickly dissipated to vibrational modes of the underlying Al lattice,
so that the formed disordered structure is not able to rearrange back
into an ordered ad-layer of oxygen atoms.
Moreover, the metal/oxygen place-exchange processes are favourable
from an electrostatic point of view, since they lead to a decrease
in the surface dipole which develops as a consequence of donation
of electronic charge from the metal to the adsorbed oxygen.
This can be quantified by computing the increase of the work function
of the metal surface upon oxidation.
This increases by 0.32~eV in the case of the spontaneously formed
oxide structure and by 0.46~eV in the case of the ordered O adlayer.
Therefore, a rearrangement of the formed disordered structure to an
ordered adlayer requires not only thermal activation to overcome the
diffusion barriers, but an additional activation energy due to the
unfavourable electrostatics.
In conclusion, we expect such rearrangements not to take
place at room temperature or below, while they are certainly not
excluded at higher oxidation temperatures (at which, however,
other ordered ultrathin oxide structures may become thermodynamically
favourable with respect to an O adlayer~\cite{Seriani_2006}).

\subsubsection*{\rm Formation of ordered oxide on TiN(001)}

The final snapshots of six consecutive FPMD simulations of
the initial oxidation of a TiN(001) surface  model
containing 9 Ti and 9 N atoms in the periodically repeated
simulation surface cell are reported in Fig.~\ref{f:o2_tin_dyns}.
As outlined above, the initial reactions of O$_2$ molecules with
the TiN(001) surface take place without incorporation of
O atoms underneath the surface.
Instead, Ti atoms are observed to escape the surface plane to become
coordinated by the incoming oxygen atoms, leaving behind Ti vacancies
in the TiN lattice (Fig.~\ref{f:o2_tin_dyns}b--e).
In this process, the surface Ti atoms leave their coordination shell
of five nitrogen atoms as a consequence of their exceedingly high
formal oxidation state when oxygen atoms are adsorbed on the 
surface~\cite{Piscanec_2004}.
The process can also be understood in terms of an increase of the
ionic character of the Ti--N bonds leading to their destabilisation after
the formation of the Ti--O bonds~\cite{Esaka_1997}.
Therefore,  the initial oxide formation on TiN appears to be
driven by a chemical driving force which induce the formation 
of a mixed-valence TiN$_{x}$O$_{y}$ surface phase~\cite{Piscanec_2004}.

In many of the simulations, dioxygen dissociation did not 
occur spontaneously, so that O$_2$ molecules remained stably
adsorbed on the partially oxidised surface for the entirety 
of the simulation runs.
In the last simulation, the temperature was gradually increased
to about 600~K within the first 2~ps of simulated time, 
then the system was annealed at this temperature for about 2~ps.
Subsequently the temperature was gradually increased to about 900~K
within 1~ps and the system annealed for about 4~ps before quenching
the atomic motion and fully minimising the atomic position.
An interesting feature observed during this simulation is the
diffusion of Ti vacancies to the second atomic layer below the
surface and the creation of Ti interstitials between the first
and second Ti layers and at the metal/oxide interface.
Moreover, the vacancies created seem to have a tendency to 
cluster together, as can be seen in Fig.~\ref{f:tin_oxide_double}
where the final structure obtained is reported showing
two periodically repeated simulation cells.
Clustering of vacancies at the metal/oxide interface has been indeed
observed experimentally upon selective Al oxidation of an
intermetallic TiAl surface~\cite{Maurice_2004}.

During the dynamics, the initially present O$_2$ molecules
dissociated and the oxide structure underwent a profound
rearrangement to form an infinite chain of fourfold-coordinated
Ti atoms (Fig.~\ref{f:tin_oxide_double}) above the existing
oxide layer.
This structure resembles very closely the proposed ``ad-molecule'' 
reconstruction of anatase TiO$_2$(001)~\cite{Lazzeri_2001}.
Other Ti atoms remained in a fivefold coordination (being bound both
to oxygen atoms and to nitrogen atoms of the substrate), as commonly
found on titanium dioxide surfaces, such as, e.g.  rutile
TiO$_2$(110)~\cite{Svetina_2001}.
The ordering achieved in the final structure is remarkable
given the short simulation time accessible to our technique.
It is probably due to the strict matching conditions imposed
by the underlying TiN lattice, on which O atoms can only bound
to Ti sites, and Ti atoms only to N sites.
Notably, at this stage it is already possible to identify
three chemically different Ti species.
Namely, (i) the fourfold-coordinated Ti atoms bound to only
oxygen, (ii) the Ti atoms at the interface bound to both
oxygen and nitrogen, and (iii) the Ti atoms of the bulk
bound to only nitrogen atoms.
These three chemical species are indeed typically found in XPS
investigations of the oxide layers grown on TiN substrates~%
\cite{Piscanec_2004,Esaka_1997,Saha_1992}.

Finally, it is to be noted that in spite of the profound
rearrangements of the oxide layer formed, the N atoms of 
the TiN lattice remained in their equilibrium positions.
Therefore, formation of thicker TiO$_{2}$ layer and release of gaseous
N$_{2}$ are expected to involve a slower, activated process.
This may be consistent with a number of XPS experiments where
TiN$_{x}$O$_{y}$ compounds have been observed to form uniformly over
the entire region of a TiN film in the early oxidation stages, prior
to the formation of an amorphous TiO$_{2}$ layer.
This is then converted into a crystalline TiO$_{2}$ film only after a
thermal treatment at higher temperatures~\cite{Esaka_1997,Saha_1992}.
Moreover, the native oxide layer formed at room temperature
has been found to be substantially thinner on TiN than on
Ti~\cite{Saha_1992}, which reflects the high stability of the
TiN$_{x}$O$_{y}$ surface phase.

\section*{5. Conclusions and outlook}

Despite evident differences among the systems investigated,
all our FPMD simulations indicate that the onset of oxide formation
on metal surfaces at low temperature is driven by the kinetic
energy released during the dissociative adsorption of O$_2$
molecules.
After the single reactions, the oxide structures obtained
become quickly frozen into a metastable state due to
energy dissipation via lattice vibration before the next
molecular reaction can take place.
In fact, the rate of collision $r$ of a molecular gas with a solid
surface is:
\begin{equation}
r = \frac{\displaystyle p}{\displaystyle \sqrt{2\pi {\rm M} k_{\rm B} {\rm T}}}\,,
\end{equation}
where M is the molecular mass, $k_{\rm B}$ the Boltzmann constant, T
the absolute temperature and $p$ the gas pressure.
From this is follows that at a pressure of 1 atmosphere and a
temperature of 300~K, 1~nm$^2$ of metal surface would be hit 
by about 3 oxygen molecules every nanosecond.
This justifies our choice of quenching the atomic motion at the end of
each simulation (lasting no more than 10~ps) and starting the
next consecutive simulation of O$_2$ adsorption from fully minimised
atomic positions.

Provided that all molecular collisions lead to dissociative oxygen
adsorption, as can be indeed observed on many metal surfaces, this
would lead to formation of a monolayer of oxide within 2 to 3~ns.
Only after formation of a few ML of oxide does the probability
of reaction with further oxygen molecules rapidly decrease, in a way
which is strongly dependent on the metal considered~\cite{Fromm_1998}.
The case of Al(111) is a peculiar one, since the fraction of oxygen
molecules which actually react upon collision (the so-called sticking
coefficient) is relatively low (10$^{-3}$ to 10$^{-2}$) at very low
oxygen coverages~\cite{Osterlund_1997}.
Therefore, an ultrathin oxide layer on Al(111) is expected to form 
in less than a microsecond under normal atmospheric conditions.

The emerging scenario is one in which surfaces exposed to an oxidising
atmosphere become almost instantaneously covered by an ultrathin oxide
layer whose structure and composition may not represent the
thermodynamical equilibrium situation.
However, this does not mean that amorphous thin oxide layers are
necessarily metastable with respect to the corresponding crystalline
phases.
As outlined in~\cite{Jeurgens_2000,Reichel_2006}, factors such as
lattice mismatch, stress in the oxide layer, formation of dislocations
at the metal/oxide interface etc.\ will have a strong influence on the
actual stability of the oxides formed.
At the level of the simulations described in this work, the
formation of vacancies and their clustering underneath the oxide layer
may have a direct influence on the stability and on the physical
properties of the native oxide layers.
This is an issue of particular importance during the formation of
native oxides on alloys via the selective oxidation of only one
element present in the alloy, and will be the subject of future
investigations.

Another factor that should be taken into account is the presence
of water vapour in the atmosphere at normal conditions, which may
condensate as a thin film on oxidised metal surfaces.
Water has been observed indeed to readily react dissociatively with an
oxidised Si(001) surface in FPMD simulations~\cite{Ciacchi_submitted}.
Thus, in general the oxide layer will be terminated by an amount
of hydroxyl groups, similarly to what is shown in the bottom
panel of Fig.~\ref{f:si_001_surf}.

Our future efforts will be directed at obtaining models of native oxide
layers on technological metal surfaces capturing their chemical
and physical properties under normal atmospheric conditions in 
a possibly realistic way.
Knowing the essential details of the structure, composition and
electronic structure of native oxide layers may then enable us 
to construct analytic potentials to study the interactions between
odixised surfaces and the external environment at the classical
simulation level, or within hybrid quantum/classical formalisms.
This will certainly be necessary in order to address processes which
occur on larger size and time scales than what is within the reach of
purely quantum mechanical schemes.
A prototypical example would be the initial adsorption on metal
surfaces of proteins which mediate the interactions between biomedical
implants and living cells, just to name a field which is of particular
interest both to the author and to whom the present journal issue is
dedicated.

\subsection*{}
The author is greatly indebted to Prof.~Wolfgang Pompe for the
precious mentoring, fruitful collaboration, and continuous support
over the last decade.
He would like to thank Prof.~Hartmut Worch for many stimulating
discussions and encouragments.
Acknowledged are long-standing collaborations with A. De Vita,
M. Stengel and M. C. Payne.
This work has been supported by the Alexander
von Humboldt Stiftung and by the Deutsche Forschungsgemeinschaft
within the Emmy-Noether Programme (CI 144/2-1).
Part of the work has been carried out under the HPC-EUROPA project
(RII3-CT-2003-506079), with the support of the European Community -
Research Infrastructure Action under the FP6 ``Structuring the
European Research Area'' Programme.
The CPU time required to perform the simulations described has been
allocated on the HPCx supercomputer facilities through the UKCP
consortium, UK, and on the supercomputers at the HLRS, Stuttgart, the
SSC, Karlsruhe and the ZIH, Dresden, Germany.
%


\vfill
\noindent 
 Correspondence address: \\
  Dr. Lucio Colombi Ciacchi \\
  Fraunhofer Institut f\"ur Werkstoffmechanik IWM, \\
  W\"ohlerstr. 11, 79098 Freiburg, Germany. \\
  Tel.: +49 761 5142113 \\
  Fax: +49 761 5142404 \\
  email: lucio@izbs.uni-karlsruhe.de

%
%

\clearpage

\begin{figure}[h!]
 \begin{center}
  \includegraphics[width=10.0cm]{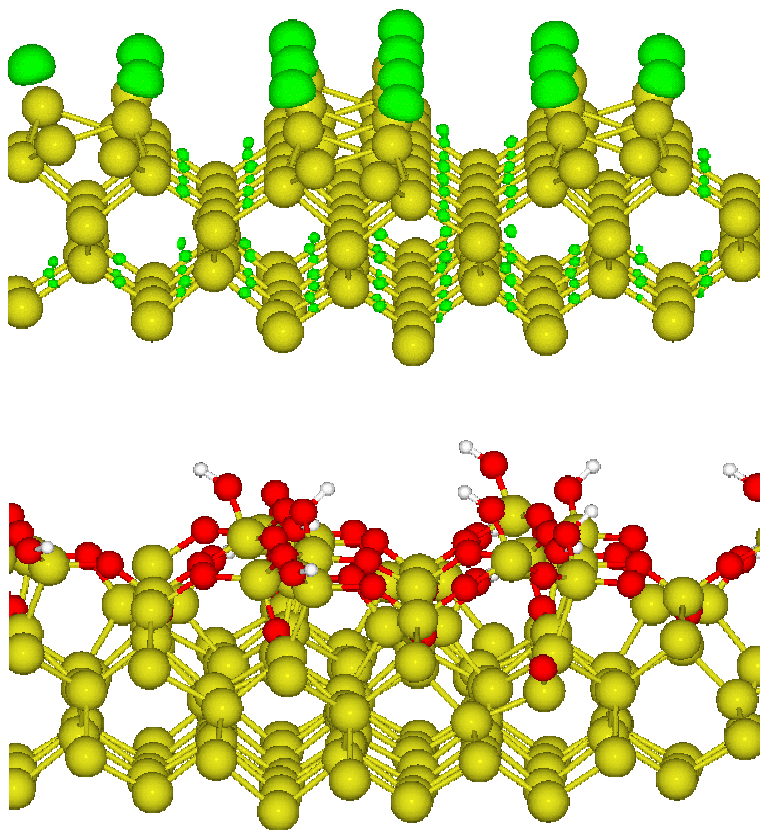}
  \caption{Top: Model of a bare reconstructed Si(001) surface
           displayed with its computed Electron Localisation Function
           (green isosurface) showing the presence of reactive
           dangling bonds on exposed Si atoms.
           Bottom: model of an oxidised and hydroxylated  Si(001)
           surface as obtained in a series of FPMD 
           simulations~\cite{Ciacchi_2005,Ciacchi_submitted}.
           \label{f:si_001_surf}}
 \end{center}
\end{figure}

\vfill 
\begin{center}
 L. Colombi Ciacchi, Figure~\ref{f:si_001_surf}.
\end{center}

\clearpage

\begin{figure}[h!]
 \begin{center}
  \includegraphics[width=15.0cm]{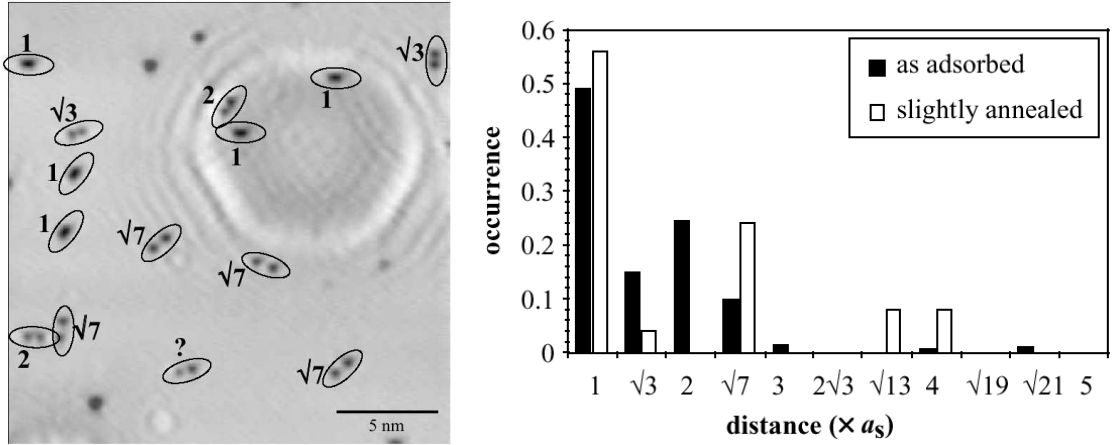}
  \caption{Left: STM image of a bare Al(111) surface upon
           exposure to an oxygen gas at 150~K. Distances
           between pairs of oxygen atoms are indicated
           in units of the Al--Al nearest neighbour distance. 
           Right: Histogram showing the relative number of
           atom pairs separated by the Al--Al distances reported 
           on the abscissa.
           The dark spots and the histogram line indicated with 
           ``1'' are in fact associated with isolated oxygen atoms 
           produced via non-adiabatic abstractive adsorption 
           processes~\cite{Komrowski_2001}.
           (Reproduced with permission from Ref.~\cite{Schmid_2001},
            copyright Elsevier 2001).
           \label{f:al_o2_stm}}
 \end{center}
\end{figure}

\vfill 
\begin{center}
 L. Colombi Ciacchi, Figure~\ref{f:al_o2_stm}.
\end{center}

\clearpage

\begin{figure}[h!]
 \begin{center}
  \includegraphics[width=15.0cm]{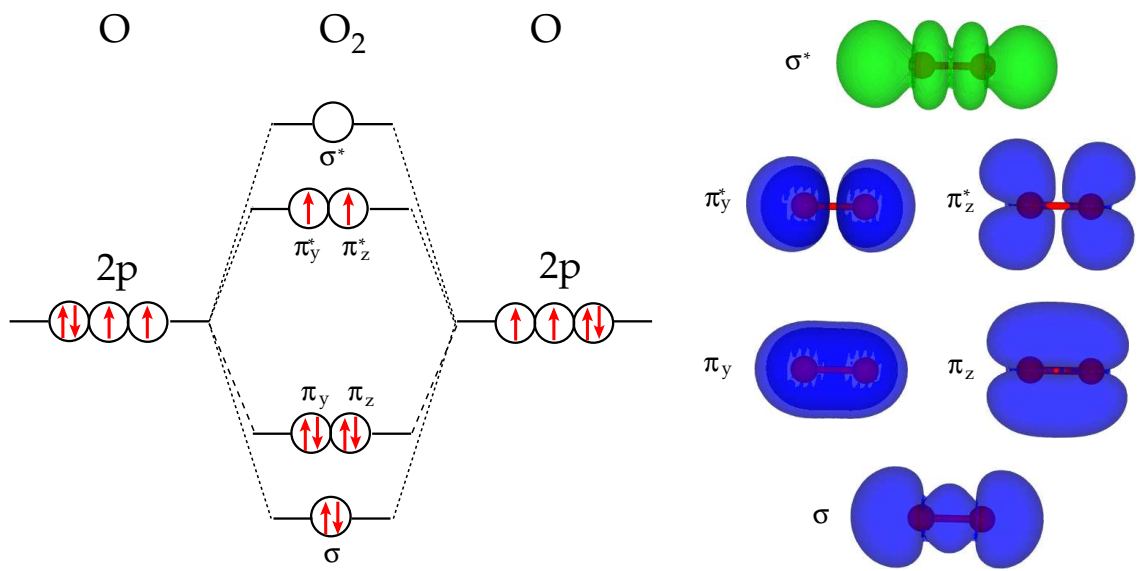}
  \caption{Scheme of the valence molecular orbitals of O$_2$
           (left) in its ground state and their corresponding 
           associated particle densities (right) represented as 
           blue and green isosurfaces for filled and empty orbitals,
           respectively. 
           \label{f:O2_orbitale}}
 \end{center}
\end{figure}

\vfill 
\begin{center}
 L. Colombi Ciacchi, Figure~\ref{f:O2_orbitale}.
\end{center}

\clearpage

\begin{figure}[h!]
 \begin{center}
  \includegraphics[width=14.0cm]{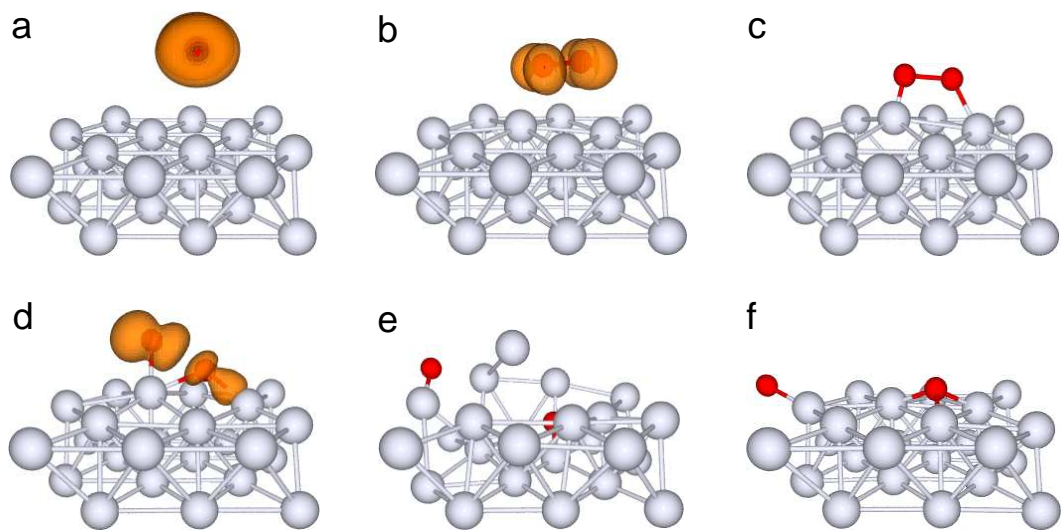}
  \caption{Snapshots from a FPMD simulation of the adsorption
           of a dioxygen molecule on the Al(111) surface~\cite{Ciacchi_2004}.
           The spin-density of the system is represented as
           an orange isosurface.
           Note the partial filling of the $\sigma^{\ast}$ antibonding
           orbital (snapshot d) which triggers the ``hot-atom''
           dissociation event.
           \label{f:al_dyn_2}}
 \end{center}
\end{figure}

\vfill 
\begin{center}
 L. Colombi Ciacchi, Figure~\ref{f:al_dyn_2}.
\end{center}

\clearpage

\begin{figure}[h!]
 \begin{center}
  \includegraphics[width=14.0cm]{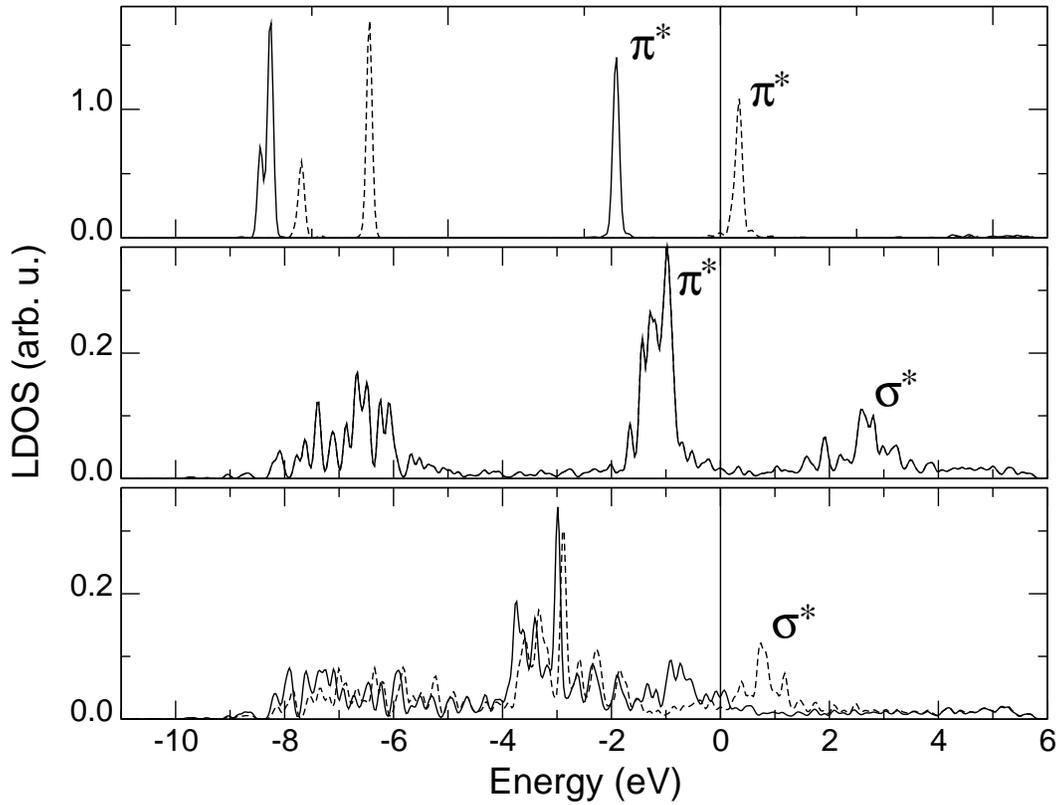}
  \caption{Evolution of the computed local density of states (LDOS)
           projected onto the O atoms during the adsorption of an
           O$_2$ molecule on the Al(111) surface.  
           The continuous and dashed lines correspond to the majority 
           and minority spin manifolds, respectively, and the Fermi 
           level is indicated with a vertical line.  
           From top to bottom, the graphs refer to an oxygen molecule 
           far from the surface, adsorbed on the surface (filling of 
           the $\pi^{\ast}$ orbitals and spin-quenching), and at the 
           initial moment of dissociation (partial filling of 
           the $\sigma^{\ast}$ orbital).
           \label{f:graph_o2_fcc_all}}
 \end{center}
\end{figure}

\vfill 
\begin{center}
 L. Colombi Ciacchi, Figure~\ref{f:graph_o2_fcc_all}.
\end{center}

\clearpage

\begin{figure}[h!]
 \begin{center}
  \includegraphics[width=14.0cm]{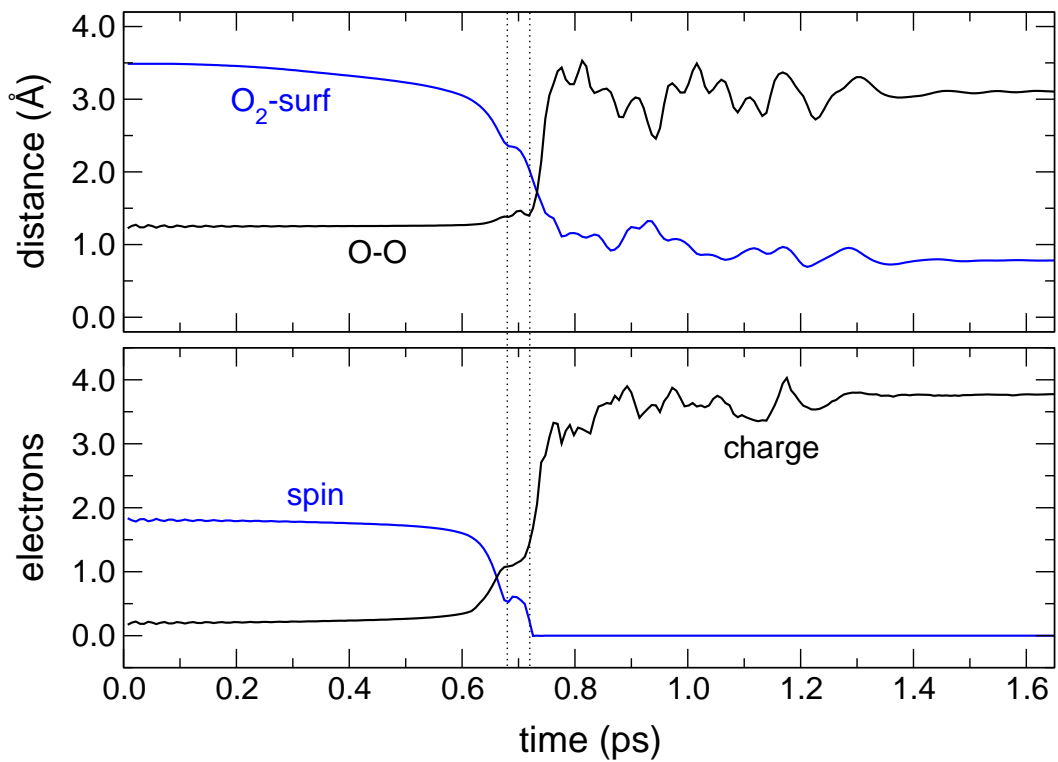}
  \caption{Dynamics of an O$_2$ molecule adsorbing dissociatively
           on the Al(100) surface simulated by FPMD.
           Top graph: evolution of the O--O distance and of the height
           of the centre of mass of the O atoms with respect to the
           average height of the surface Al atoms.
           Bottom graph: evolution of the spin-density and of the 
           atomic charge integrated within the Bader regions 
           associated with both O atoms.
           \label{f:graph_o2_100_dist_char}}
 \end{center}
\end{figure}

\vfill 
\begin{center}
 L. Colombi Ciacchi, Figure~\ref{f:graph_o2_100_dist_char}.
\end{center}

\clearpage

\begin{figure}[h!]
 \begin{center}
  \includegraphics[width=16.0cm]{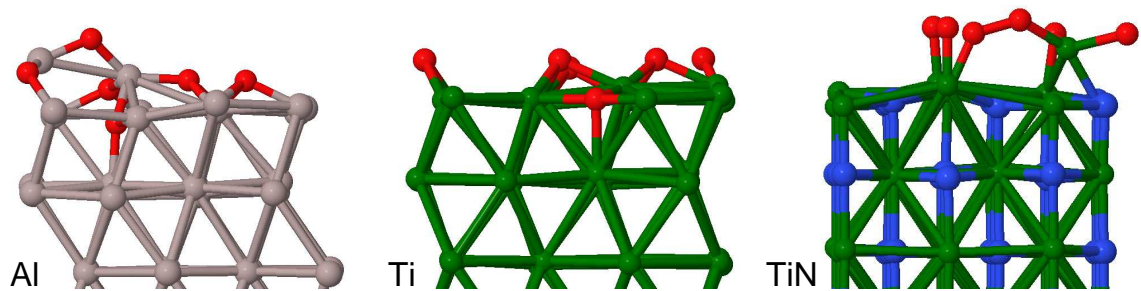}
  \caption{Structures of partially oxidised metal surfaces
           as obtained in FPMD simulations (only the three 
           topmost metal layers are shown, for clarity). 
           The oxygen atom coverage of the surfaces is 0.5 
           in the case of Al(111) and Ti(111) 
           and 0.67 in the case of TiN(001).
           Note the spontaneous incorporation of O atoms underneath
           the surface in the case of Al and Ti, the extraction
           of metal atoms from the surface layer in the case of
           Al and TiN, and the presence of an adsorbed O$_2$ 
           molecule on TiN.
           \label{f:o2_03_surfs}}
 \end{center}
\end{figure}

\vfill 
\begin{center}
 L. Colombi Ciacchi, Figure~\ref{f:o2_03_surfs}.
\end{center}

\clearpage

\begin{figure}[h!]
 \begin{center}
  \includegraphics[width=12.0cm]{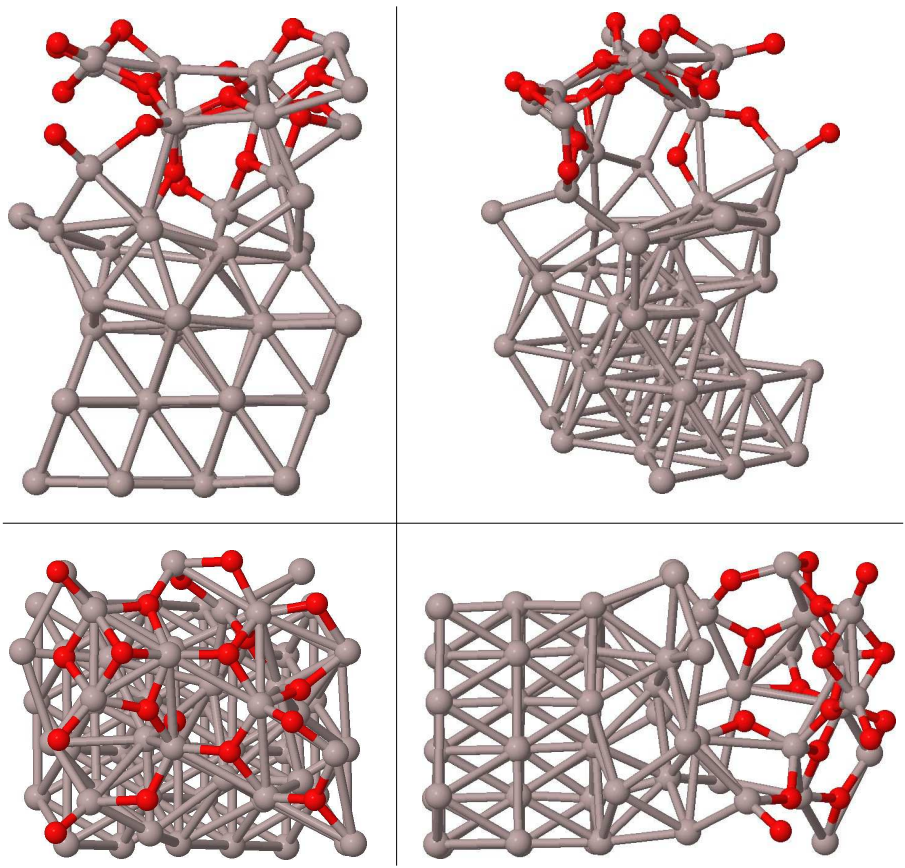}
  \caption{Model of a native oxide layer grown on the Al(111) 
           surface (the oxygen coverage is 1.5~ML), as obtained in a series 
           of FPMD simulations. Orthogonal projections and
           perspective view (top right).
           \label{f:oxide_1.5ML_al_111}}
 \end{center}
\end{figure}

\vfill 
\begin{center}
 L. Colombi Ciacchi, Figure~\ref{f:oxide_1.5ML_al_111}.
\end{center}

\clearpage

\begin{figure}[h!]
 \begin{center}
  \includegraphics[width=13.0cm]{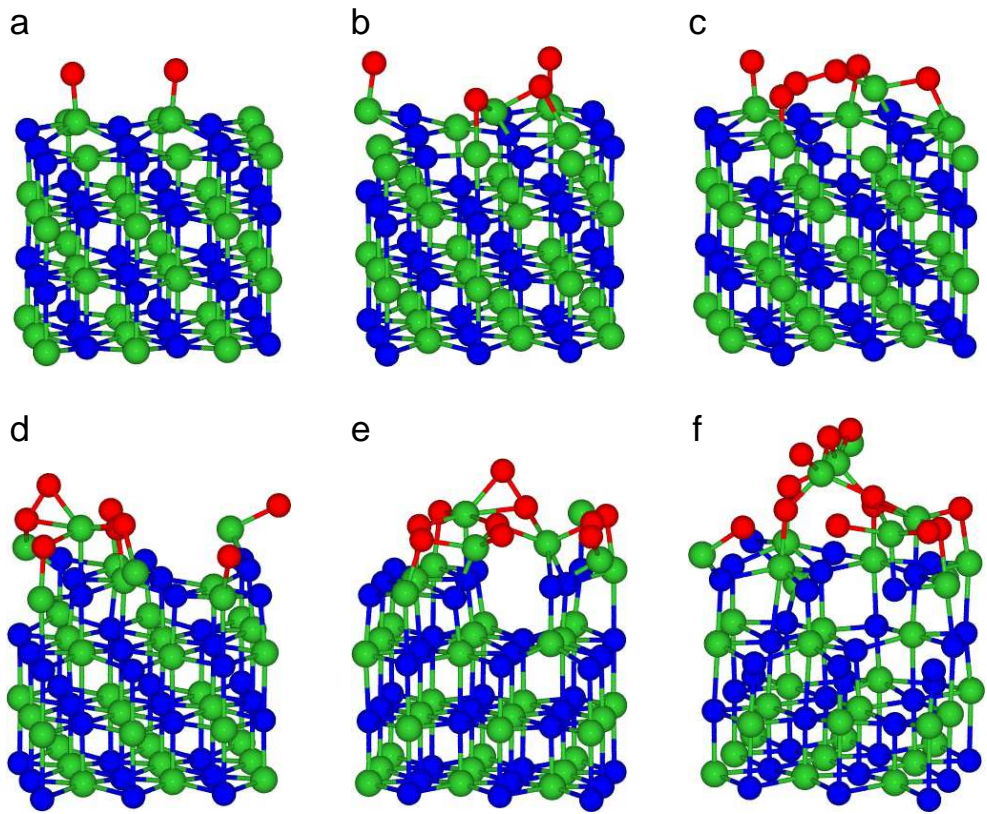}
  \caption{Final snapshots of consecutive FPMD simulations
           of the initial oxide formation on the TiN(001)
           surface. 
           The oxygen covrage increases from (a) 0.22~ML
           to (f) 1.33~ML. 
           \label{f:o2_tin_dyns}}
 \end{center}
\end{figure}

\vfill 
\begin{center}
 L. Colombi Ciacchi, Figure~\ref{f:o2_tin_dyns}.
\end{center}

\clearpage

\begin{figure}[h!]
 \begin{center}
  \includegraphics[width=13.0cm]{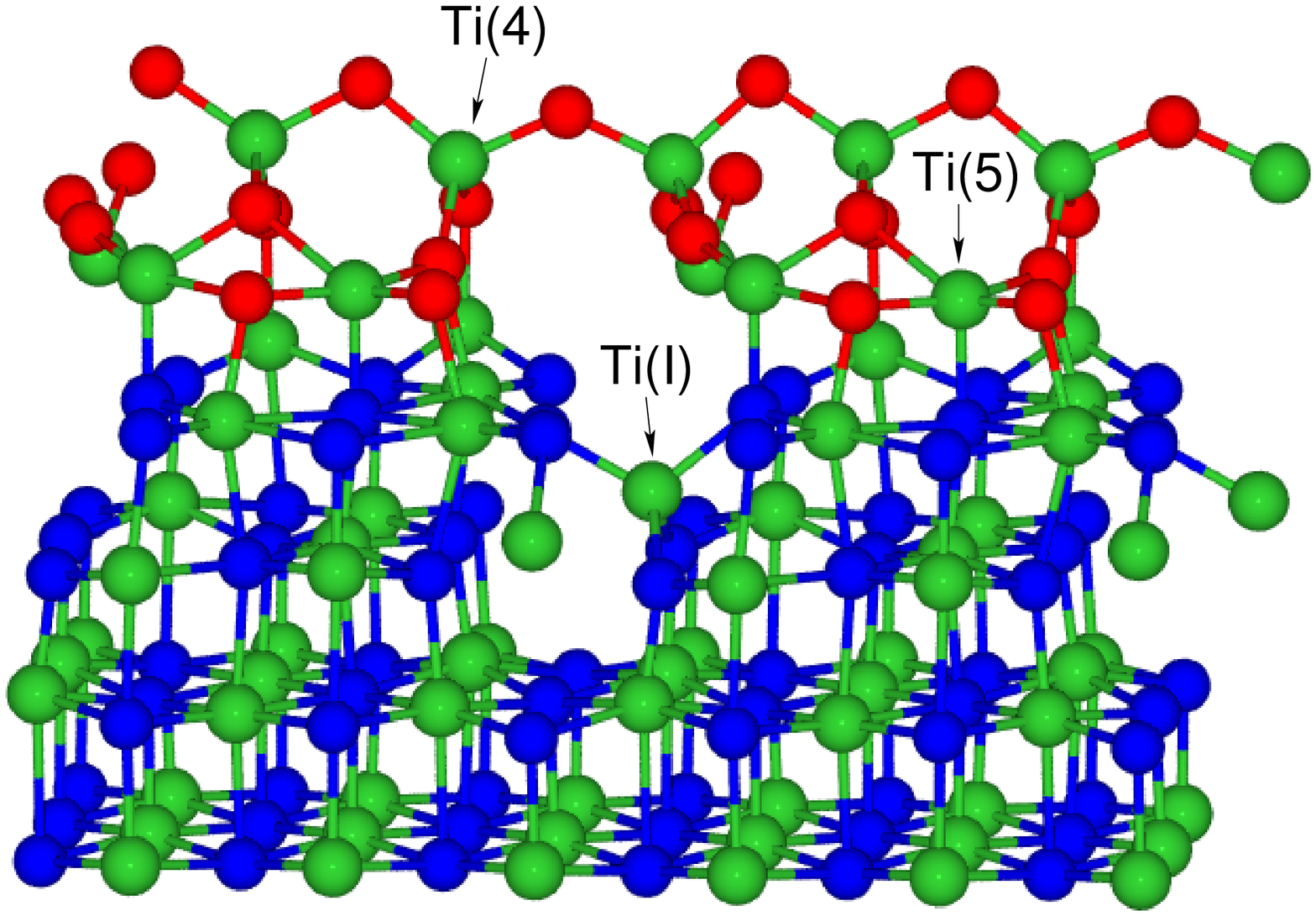}
  \caption{Model of a native oxide layer grown on the TiN(001)
           surface (the oxygen coverage is 1.33~ML), as obtained in a series 
           of FPMD simulations.
           A fourfold-coordinated Ti atom is indicated with Ti(4),
           a fivefold-coordinated Ti atom with Ti(5), and an
           atom in interstitial position with respect to the
           TiN lattice with Ti(I).
           \label{f:tin_oxide_double}}
 \end{center}
\end{figure}

\vfill 
\begin{center}
 L. Colombi Ciacchi, Figure~\ref{f:tin_oxide_double}.
\end{center}


\begin{thebibliography}{99}

\bibitem{Lawless_1974}
 K. R. Lawless: Rep. Prog. Phys. 37 (1974) 231-316.

\bibitem{Hua_2006}
Z. Zhen-Hua, D. Hui-Qiu, L. Wei-Xue, H. Wang-Yu:
Acta Phys. Sin. 55 (2006) 3157-3164.

\bibitem{Piscanec_2004}
 S. Piscanec, L. Colombi Ciacchi, E. Vesselli, G. Comelli, O. Sbaizero, 
 S. Meriani, A. De Vita: Acta Mater. 52 (2004) 1237-1245.

\bibitem{Jerkiewicz_2004}
 G. Jerkiewicz, G. Vatankhah, J. Lessard, M. P. Soriaga, Y.-S. Park:
 Electrochim. Acta 49 (2004) 1451-1459.

\bibitem{Diniz_2004}
 F. B. Diniz, R. R. Ueta: Electrochim. Acta 49 (2004) 4281-4286.

\bibitem{Fehlner_1986}
 F. P. Fehlner, in: Low temperature oxidation, John Wiley and Sons, 
 New York (1986).

\bibitem{Muhlstein_2002}
 C. L. Muhlstein, E. A. Stach, R. O. Ritchie: Acta Mater. 50 (200) 3579-3595.

\bibitem{Fromm_1998}
 E. Fromm, in: Kinetics of metal-gas interactions at low temperatures,
 Springer-Verlag, Berlin Heidelberg (1998).

\bibitem{Tammann_1920}
 G. Tamann: Z. Anorg. Allgem. Chem. 111 (1920) 78-89.

\bibitem{Cabrera_1948}
 N. Cabrera, N. F. Mott: Rep. Prog. Phys. 12 (1948) 163-184.

\bibitem{Fehlner_1970}
 F. P. Fehlner, N. F. Mott: Oxid. Met. 2 (1970) 59-99.

\bibitem{Jeurgens_2002}
L. P. H. Jeurgens, W. G. Sloof, F. D. Tichelaar, E. J. Mittemeijer:
J. Appl. Phys. 92 (2002) 1649-1656.

\bibitem{Reichel_2006}
F. Reichel, L. P. H. Jeurgens, E. J. Mittemeijer:
Phys. Rev. B 74 (2006) 144103.

\bibitem{Lundgren_2002}
 E. Lundgren, G.  Kresse, C. Klein, M. Borg, J. N. Andersen, 
 M. De Santis, Y. Gauthier, C. Konvicka, M. Schmid, P. Varga:
 Phys. Rev. Lett. 88 (2002) 246103.

\bibitem{Benedek_1999}
 R. Benedek, D. N. Seidman, M. Minkoff, L. H. Yang, A. Alavi:
Phys. Rev. B 60 (1999) 16094-16102.

\bibitem{Gemming_2001}
S. K\"ostlmeier-Gemming, C. Els\"asser:
Phys. Chem. Chem. Phys. 3 (2001) 5140-5144.

\bibitem{Ruckman_1992}
 M. W. Ruckman, J. Chen, M. Strongin, E. Horache:
 Phys. Rev. B 45 (1992) 14273-14278.

\bibitem{Jacobsen_1995}
 F. M. Jacobsen, S. Raaen, M. W. Ruckman, M. Strongin:
 Phys. Rev. B 52 (1995) 11339-11342.

\bibitem{Payne_1992}
 M. C. Payne, M. P. Teter, D. C. Allan, T. A. Arias, J. D. Joannopoulos:
 Rev. Mod. Phys. 64 (1992) 1045-1097.

\bibitem{Hafner_2000}
 J. Hafner: Acta Mater. 48 (2000) 71-92.

\bibitem{Reuter_2003}
 K. Reuter, M. Scheffler: Phys. Rev. Lett. 90 (2003) 046103.

\bibitem{Finnis_2005}
 M. W. Finnis, A. Y. Lozovoi, A. Alavi:
 Annu. Rev. Mater. Res. 35 (2005) 167-207.

\bibitem{Vashishta_1999}
T. Campbell, R. K. Kalia, A. Nakano, P. Vashishta, S. Ogata, S. Rodgers:
Phys. Rev. Lett. 82 (1999) 4866-4869.

\bibitem{Vashishta_2006}
P. Vashishta, R. K. Kalia, A. Nakano:
J. Phys. Chem. B 110 (2006) 3727-3733.

\bibitem{Galli_1993}
 G. Galli, A. Pasquarello, in: M. P. Allen, D. J. Tildesley (Eds.),
 Computer Simulation in Chemical Physics, Kluwer Academic Publishers,
 Dordrecht (1993) 261-313.

\bibitem{Marx_2000}
 D. Marx, J. Hutter, in: J. Grotendorst (Ed.), Modern Methods 
 and Algorithms of Quantum Chemistry, John von Neumann
 Institute of Computing, J\"ulich (2000) 301-449.

\bibitem{Verlet_1967}
 L. Verlet: Phys. Rev. 156 (1967) 98-103.

\bibitem{Lautrec}
 A. De Vita, A. Canning, G. Galli, F. Gygi, F. Mauri, R. Car: 
 EPFL Supercomput. Rev. 6 (1994) 22.

\bibitem{Car_1985}
 R. Car,  M. Parrinello: Phys. Rev. Lett. 55 (1985) 2471-2474.

\bibitem{Perdew_1992}
 J. P. Perdew, Y. Wang: Phys. Rev. B, 45 (1992) 13244-13249.

\bibitem{Troullier_1993}
 N. Troullier, J. L. Martins: Phys. Rev. B 43 (1991) 1993-2006.

\bibitem{Blochl_1994}
 P. E. Bl\"ochl: Phys. Rev. B 50 (1994) 17953-17979. 

\bibitem{VandeVondele_1999}
 J. VandeVondele, A. De Vita: Phys. Rev. B 60 (1999) 13241-13244.

\bibitem{Stengel_2000}
 M. Stengel, A. De Vita: Phys. Rev. B 62 (2000) 15283-15286.

\bibitem{Ciacchi_2004}
 L. C. Ciacchi, M. C. Payne: Phys. Rev. Lett. 92 (2004) 176104.

\bibitem{Ciacchi_2005}
 L. C. Ciacchi, M. C. Payne: Phys. Rev. Lett. 95 (2005) 196101.

\bibitem{Becke_1990}
 A. D. Becke, K. E. Edgecombe: J. Chem. Phys. 92 (1990) 5397-5403.

\bibitem{Brune_1992}
 H. Brune, J. Wintterlin, R. J. Behm, G. Ertl: 
 Phys. Rev. Lett 68 (1992) 624-627.

\bibitem{Schmid_2001}
 M. Schmid, G. Leonardelli, R. Tschelie\ss nig, A. Biedermann, P. Varga:
 Surf. Sci. 478 (2001) L355-L-362.

\bibitem{Komrowski_2001}
 A. J. Komrowski, J. Z. Sexton, A. C. Kummel:
 Phys. Rev. Lett. 87 (2001) 246103.

\bibitem{Wintterlin_1996}
 J. Wintterlin, R. Schuster, G. Ertl:
 Phys. Rev. Lett. 77 (1996) 123-126.

\bibitem{Yourd_2002}
Y. Yourdshahyan, B. Razaznejad, B. I. Lundqvist:
Phys. Rev. B 65 (2002) 075416.

\bibitem{Bader_1990}
 R. F. W. Bader, in: Atoms in Molecules: A Quantum Theory, 
 Oxford University Press, Oxford (1990).

\bibitem{Brune_1993}
 H. Brune, J. Wintterlin, G. Ertl, J. Wiechers, R. J. Behm:
 J. Chem. Phys. 99 (1993) 2128-2148.

\bibitem{Snijders_2005}
 P. C. Snijders, L. P. H. Jeurgens, W. G. Sloof:
 Surf. Sci. 589 (2005) 98-105.

\bibitem{Seriani_2006}
 N. Seriani, W. Pompe, L. C. Ciacchi:
 J. Phys. Chem. B 110 (2006) 14860-14869.

\bibitem{Esaka_1997}
 F. Esaka, K. Furuya, H. Shimada, M. Imamura, N. Matsubayashi, 
 H. Sato, A. Nishijima, A. Kawana, H. Ichimura, T. Kikuchi:
 J. Vac. Sci. Technol. A 15 (1997) 2521-2528.

\bibitem{Maurice_2004}
 V. Maurice, G. Despert, S. Zanna, M.-P. Bacos, P. Marcus:
 Nature Mater. 3 (2004) 687-691.

\bibitem{Lazzeri_2001}
 M. Lazzeri, A. Selloni: Phys. Rev. Lett. 287 (2001) 266105.

\bibitem{Svetina_2001}
 M. Svetina, L. Colombi Ciacchi, O. Sbaizero, S. Meriani, A. De Vita:
 Acta Mater. 49 (2001) 2169-2177.

\bibitem{Saha_1992}
 N. C. Saha, J. Tompkins: J. Appl. Phys. 72 (1992) 3072-3079.

\bibitem{Osterlund_1997}
 L. \"Osterlund, I. Zori\'c, B. Kasemo:
 Phys. Rev. B 55 (1997) 15452-15455.

\bibitem{Jeurgens_2000}
 L. P. H. Jeurgens, W. G. Sloof, F. D. Tichelaar, E. J. Mittemeijer:
 Phys. Rev. B 62 (2000) 4707-4719.

\bibitem{Ciacchi_submitted}
 L. C. Ciacchi, J. Bagdahn, D. J. Cole, M. C. Payne, P. Gumbsch,
 submitted for publication.


\end{thebibliography}
\end{document}